\documentclass[man, 12pt, floatsintext]{apa7}

\usepackage[american]{babel}
\usepackage[T1]{fontenc}
\usepackage[style=apa, citestyle=authoryear-comp, uniquelist=false, sorting=nyt, backend=biber, uniquename=false, hyperref=true, sortcites=false]{biblatex}
\usepackage{mathptmx}
\usepackage{url}
\usepackage{amsmath}
\usepackage{amssymb}
\usepackage{caption}
\usepackage{graphicx}
\usepackage{tikz}
\usepackage{orcidlink}
\usepackage[frozencache=true,cachedir=minted-cache]{minted} 
\usepackage{enumitem}
\usepackage{makecell}
\usepackage{adjustbox}
\usepackage{multirow}
\usepackage{mathtools}
\usepackage{csquotes}
\usepackage{xfrac}
\usepackage{bm}
\usepackage{setspace}
\usepackage{listings}
\usepackage{tabularx}

\DeclareLanguageMapping{american}{american-apa}
\DeclareUnicodeCharacter{0302}{\^{R}}
\DeclareDelimFormat{nameyeardelim}{\addcomma\space}
\DeclareMathAlphabet{\pazocal}{OMS}{zplm}{m}{n}
\SetMathAlphabet\pazocal{bold}{OMS}{zplm}{bx}{n}
\definecolor{beige}{HTML}{FAF7F3}
\definecolor{lightblue}{HTML}{E2EBF2}
\definecolor{midblue}{HTML}{BFD3E2}
\definecolor{navyblue}{HTML}{5784AA}
\definecolor{blueberryblue}{HTML}{0041C2}

\title{Bayesian Multilevel Compositional Data Analysis with the R Package \textbf{\textit{multilevelcoda}}}
\shorttitle{multilevelcoda R package}
\author{
Flora Le\orcidlink{0000-0003-0089-8167}$^1$, 
Dorothea Dumuid\orcidlink{0000-0003-3057-0963}$^2$, 
Tyman E. Stanford\orcidlink{0000-0002-8570-5493}$^2$, 
Joshua F. Wiley\orcidlink{0000-0002-0271-6702}$^1$
}
\authorsaffiliations{
$^1$School of Psychological Sciences{,} ,
Monash University
\\
$^2$Alliance for Research in Exercise{,} Nutrition and Activity{,},
Allied Health and Human Performance{,} ,
University of South Australia
}
\authornote{
Reproducible materials for this study are available at: \url{https://github.com/florale/multilevelcoda-overview}.
\\
FL: Conceptualization, formal analysis, investigation, methodology, project administration, software, visualization, writing-original draft, writing-review and editing.
DD: Conceptualization, resources, supervision, writing-review and editing.
TES: Conceptualization, resources, supervision, writing-review and editing.
JFW: Conceptualization, investigation, methodology, resources, software, supervision, writing-review and editing.
\\ 
FL was supported by a Monash Graduate Scholarship and a Monash International Tuition Scholarship. DD was funded by an Australian Research Council (ARC) Discovery Early Career Award (DECRA) DE230101174 and by the Centre of Research Excellence in Driving Global Investment in Adolescent Health funded by National Health and Medical Research Council (NHMRC) GNT1171981. JFW was supported by a NHMRC fellowship (1178487). The authors declared no conflict of interests. 
\\
Correspondence to 
Flora Le (\url{flora.le@monash.edu}) or 
Joshua F. Wiley (\url{joshua.wiley@monash.edu}),
School of Psychological Sciences, 
Monash University, Clayton, VIC, Australia.
}
\leftheader{Le, Dumuid, Stanford, and Wiley}

\abstract{
Multilevel compositional data, such as data sampled over time that
are non-negative and sum to a constant value, are common in various
fields. However, there is currently no software specifically built to
model compositional data in a multilevel framework. The
\textbf{R} package \textbf{\textit{multilevelcoda}} implements a collection of
tools for modelling compositional data in a Bayesian multivariate,
multilevel pipeline. The user-friendly setup only requires the data,
model formula, and minimal specification of the analysis. This paper
outlines the statistical theory underlying the Bayesian compositional
multilevel modelling approach and details the implementation of the
functions available in \textbf{\textit{multilevelcoda}}, using an example dataset of
compositional daily sleep-wake behaviours. 
This innovative method can be used to gain robust answers to scientific questions using the increasingly available multilevel compositional data 
from intensive, longitudinal studies. 
}

\begin{document}
\maketitle
Bayesian approaches have been increasingly employed for multilevel models. 
Motivations for using Bayesian approaches have
been covered extensively in other work, including flexibility to specify complex
models \parencite{levy2023} like non-normal random-effect models, robustness
to small sample sizes \parencite{le2024, stegmueller2013}, benefits from
incorporating existing empirical information
\parencite[i.e., priors;][]{van2018}, and the ease of quantifying
uncertainty around arbitrary calculated quantities using posterior
samples \parencite{gelman2013bayesian, wagenmakers2016}.
Bayesian sampling algorithms, 
including the Markov chain Monte Carlo (MCMC) sampling, or
Hamiltonian Monte Carlo (HMC) \parencite{betancourt2017, betancourt2014} and
its extension, the No-U-Turn Sampler (NUTS) \parencite{hoffman2014}. MCMC
allows for both model parameter estimation and drawing samples from the
posterior predictive distribution, which can be used to assess model
fit. Importantly, the individual posterior distribution samples can be
used to make inferences about the parameters, such as calculating the
mean, standard errors and credible intervals, enabling post-hoc analyses
involving any calculated quantities to be directly and intuitively
conducted from Bayesian multilevel models.
Such features can greatly facilitate the analysis
of data with complex structure, 
such as multilevel compositional data.

Multilevel compositional data exist in various fields, 
such as time-use epidemiology 
(e.g., time spent in different sleep-wake behaviours during the 24-hour day), 
and sleep (e.g., proportion of time spent in different sleep stages during the night),
and nutritional epidemiology (e.g., macronutrients like proteins, fats and
carbohydrates as proportions of total caloric intake). 
These data can be classified as compositions, 
which consist of \textit{compositional parts} that contain
relative information about the whole; represented as non-negative values
that sum to a constant value. Compositional parts can be expressed as
percentages (or proportions) of the composition but also may be in other
units that are constrained to a constant total value (e.g., 1440 minutes
in a day). 
They are commonly
measured across multiple time points (e.g., across several consecutive
days), or nested within clusters (e.g., schools).
This means the data are multilevel, 
with the most common multilevel data structure having two levels 
(e.g., consecutive days nested within people). 
Thus, these data often consist of two sources of
variability at each level: between
(differences between clusters, such as people) and within (differences
within clusters, commonly the deviation of a specific value from the
average of that cluster).

Despite the abundance of multilevel compositional data, 
standard statistical methods, including multilevel models, do not produce 
valid results for raw compositional data.
This is due to the 
perfect multicollinearity present in their constant-sum nature (i.e.,
compositional parts are linearly dependent).
Instead, compositional data analysis
\parencite[CoDA;][]{Aitchison1986, Pawlowsky2011} 
utilises the relative information contained in
compositional data using log-ratios. 
Certain log-ratio transformations
can remove the linear dependence of compositional parts, while retaining the relative
nature of the compositional components (i.e., changes in compositional
components but not their total).
One such transform is the isometric log-ratio tranform (ilr), 
that is the most commonly used in the 
physical activity and sedentary behaviour research.
The ilr transformation eliminates the multicollinearity by producing
one less ilr coordinate compared to the number of compositional parts 
(e.g. two ilr coordinates are calculated from a 3-part composition),
thus allowing standard statistical methods to be applied on the transformed data.
Some software for compositional data analysis exist, mostly in
\textbf{R}, including \textbf{\textit{compositions}}
\parencite{compositions, van2013}, \textbf{\textit{compositions}}
\parencite{zCompositions}, \textbf{\textit{robCompositions}} \parencite{robcompositions},
\textbf{\textit{Compositional}} \parencite{Compositional}, \textbf{\textit{codaredistlm}}
\parencite{dumuid2018, codaredistlm}. 
These packages offer general tools for manipulating or
modelling compositional data so they can be used in standard statistical
models. However, they do not easily accommodate the manipulation of
compositional data with a multilevel structure or provide functions to
fit multilevel models with compositional data 
(or their corresponding log-ratios) 
as response or predictor variables.

Further, to facilitate the interpretability of CoDA, 
isotemporal compositional
substitution analysis \parencite{Dumuid2019} is a post-hoc approach to
examine the model predicted changes in an outcome associated with
changes to the compositional parts. Substitution analysis provides an opportunity
to interrogate the model to answer questions about the predicted change
in an outcome when the compositional parts are redistributed.
This analysis can answer questions such as
whether there is a change in health when people spend
time being physically active at the expense of sitting. Increasing
evidence from isotemporal compositional substitution analysis shows
that reallocating time between behaviours are associated with both physical
and mental heath outcomes \parencite{janssen2020, grgic2018, miatke2023}.
Most existing studies are, however, cross-sectional, with less evidence
available from longitudinal data. This may be due to the lack of tools
to efficiently work with longitudinal compositional data in substitution analysis. 
To our knowledge, no software
currently automates isotemporal compositional
substitution analysis, especially in a multilevel framework.

The \textbf{\textit{multilevelcoda}} package \parencite{multilevelcoda}, presented in this article, 
aims to address these gaps with tools to automate estimating multilevel
models for compositional data (and their associated log-ratios) and isotemporal compositional 
substitution analysis in a Bayesian framework. 
Specifically, 
\textbf{\textit{multilevelcoda}} advances the analysis of multilevel compositional data
by offering three important contributions:

\begin{itemize}
\item
  Compute multilevel compositions and perform log-ratio transformation.
  Decompose data into between and within levels if necessary.
\item
  Automatically fit Bayesian multilevel models with compositional
  predictors and/or outcomes. The Bayesian models are implemented using
  the \textbf{\textit{brms}} package, which supports a variety of generalised
  (non-)linear multivariate multilevel models.
\item
  Estimate isotemporal compositional substitution analysis for Bayesian
  multilevel models at both between and within levels. We leverage the
  posterior draws of Bayesian models to derive reliable estimates of
  credible intervals of the predicted differences in
  the expected outcome for the reallocation of compositional parts.
\end{itemize}
We begin by introducing the fundamental concepts of multilevel
composition and the underlying multilevel models for compositional data.
We then describe the functionality
of the software using an example of daily 24-hour sleep-wake behaviours. 
Lastly, we provide a comparison across packages for
compositional data analysis and discuss plans for extending the package.

\section{Multilevel Composition Data}
\subsection{Properties of Compositional Data}
\label{logratio}

A \textit{composition} is defined as a vector of \(D\) positive
components, called \textit{compositional parts}, that sum to a constant
\(\kappa\) 

\begin{equation}
\pmb{x} = (x_{1}, x_{2}, \ldots, x_{D}), \hspace{0.1cm}
\end{equation}
where  $\sum_{i=1}^D x_i = \kappa$ 
and $x_i > 0 \hspace{0.2cm} \forall i = 1,2,\ldots, D$.
Consequently, compositions are elements in the \(D\)-simplex, denoted as
\(\mathcal{S}^D \subset \mathbb{R}^D\), whose parts are linearly
dependent as each part can be deduced with knowledge of the \(D-1\)
other parts (e.g., \(x_j = \kappa - \sum_{\forall i \neq j} x_i\)). 
An important operation on the simplex is perturbation
\parencite{Aitchison1986}, or closure operation applied to the element-wise
product. Perturbation is the analogous operation to addition in the
Euclidean space \parencite{van2013, Aitchison1986}, defined as

\begin{equation}
\pmb{x} \oplus \pmb{x^{*}} = \mathcal{C}(x_1 
\pmb{\cdot} x^{*}_1, 
x_2 \pmb{\cdot} x^{*}_2, 
\ldots, 
x_D \pmb{\cdot} x^{*}_D)
\end{equation}
where
\[\mathcal{C}\left( \pmb{x}\right)= \frac{\kappa}{\sum_{i=1}^D x_i}\pmb{x}\]
is the closure operation that normalises the compositional parts of a
vector \(\pmb{x}\) to the constant \(\kappa\)
\parencite{Aitchison1986}, and
\(\pmb{x}, \pmb{x^{*}} \subset \mathcal{S}^D\). Such
properties are incompatible with many standard mathematical operations
(e.g., \(\mathcal{S}^D\) is not closed under addition) and statistical
methods that assume independence (e.g., multiple linear regression) that
are developed in the real space \(\mathbb{R}^{D - 1}\)
\parencite[for detailed discussion on the properties of compositional data and their consequences, see][]{Aitchison1994, Aitchison1986}.

\subsection {Log-ratio Approach for Multilevel Compositional Data Analysis}
Two modelling strategies can be used for compositional data: a) directly
working on the simplex, and b) transforming compositional data from the
simplex to the real space then modelling the transformed data using a
multivariate normal distribution, which is referred to as principle of
working in coordinates \parencite{mateu2011}. The most widely studied
distribution on the simplex is the Dirichlet, which can be used to model
composition directly \parencite{gueorguieva2008}. However, its use in
applications is quite limited, as the strong independence structure of
Dirichlet distribution (i.e., independent, equally scaled
gamma-distributed variables) poorly models the dependence between
compositional components \parencite{aitchison1982}.

The alternative approach to modelling compositional data, originating
from \cite{aitchison1982}, focuses on the relative magnitudes and
variations of components, rather than their absolute values. The
prototype of a distribution in the simplex is the logistic-normal
distribution
\parencite[additive log-ratio transformation;][]{atchison1980, aitchison1982},
which is also referred to as normal distribution on the simplex. Using
log-ratios, a composition in the simplex (\(\mathcal{S}^D\)) can be
expressed in terms of ratios of the components of the composition in the
Euclidean space (\(\mathbb{R}^{D - 1}\)) where standard mathematical
operations and statistical methods are valid. A family of log-ratio
transformations for modelling compositional data includes additive
log-ratio \parencite[alr;][]{aitchison1982}, centered log-ratio
\parencite[clr;][]{aitchison1982}, and isometric log-ratio
\parencite[ilr;][]{Egozcue2003}.
Non-orthonomal transformations (e.g., alr, clr, or simple log-ratio
transformations) 
have limitations in modelling compositional data
\parencite{mateu2011}. Specifically,
the alr transformation is not isometric, thus, 
does not preserve the properties of
compositional data (angles and distances). 
The clr transformation preserves the distance but does not break
the sum constraint, which results in a singular covariance matrix.
In contrast, the ilr-transform maps the \(D\)-part compositional data from the simplex to non-overlapping subgroups in the \((D - 1)\)-dimension
Euclidean space isometrically by using an orthonormal basis, thereby
preserving the compositional properties and yielding a full-rank
covariance matrix. We describe ilr transform in detail in the following.

Consider a composition
\(\pmb{x} \in \mathcal{S}^D\) and a corresponding set of
\((D - 1)\) ilr coordinates
\((z_1, z_2,\ldots,z_{D-1} ) = \pmb{z}  \in \mathbb{R}^{D - 1}\).
The individual \(z_k\) coordinate is constructed as normalised log-ratio
of the geometric mean of compositional parts in the numerator (a
mutually exclusive set of subcompositions denoted as \(R_{k}\)) to the
geometric mean of compositional parts in the denominator (a set of
subcompositions denoted as \(S_{k}\)). The \(k^{\text{th}}\)
(\(k = 1, 2,\ldots, D-1\)) ilr coordinate can then be written as

\begin{equation}
z_{k} 
=
\sqrt{\frac{r_{k}s_{k}}{r_{k}\ +s_{k}}}
\text{ln}\left(\frac{\tilde{x}_{R_{k}}}{\tilde{x}_{S_{k}}}\right), 
\hspace{0.2cm} k = 1, 2, \ldots, D - 1
\end{equation}
where \[
\tilde{x}_{R_{k}} = \left(\prod_{x_{d} \in R_{k}}x_{d}\right)^\frac{1}{r_{k}}
\hspace{0.2cm} \text{and} \hspace{0.2cm}
\tilde{x}_{S_{k}} = \left(\prod_{x_{d} \in S_{k}}x_{d} \right)^\frac{1}{s_{k}}
\]
with \(r_k\) and \(s_k\) being the size of the sets \(R_k\) and \(S_k\),
respectively, and \(\sqrt{\frac{r_{k}s_{k}}{r_{k}\ + s_{k}}}\) being a
normalising constant.

One method for ilr transformation employs a sequential binary partition
(SBP) process \parencite{egozcue2005}, which produces ilr coordinates that
are interpretable depending on the application. A SBP is
obtained by first partitioning the compositional parts into two
non-empty sets, where one set corresponds to the first ilr coordinate's
numerator and the other set corresponds to the first ilr coordinate's
denominator. Using the same principle, each of the previously
constructed sets are recursively partitioned into two non-empty sets
until no further non-empty partitions of the subcompositional parts are
possible (after \(D - 1\) steps). This SBP process can be coded via a
\(D \times (D-1)\) matrix corresponding to the \(D\) compositional parts
and their membership in the \((D-1)\) ilr coordinates; +1 if the
compositional part is the ilr numerator, -1 if the compositional part is
the ilr denominator, or 0 if the compositional part is uninvolved in the
ilr coordinate. The ilr coordinates can be interpreted as the log ratio
of the subcomposition in the numerator in relation to the subcomposition
in the denominator.

SBP can be constructed to form conceptually meaningful contrasts
(e.g., time spent in sleeping behaviours all relative to waking behaviours).
In some cases, it is not straightforward to correctly interpret ilr
coordinates, as they are expressed in terms of the log-ratios of groups
of parts. Another choice of SBP for ilr transformation involves
constructing the ilr coordinate \(z_k\) to capture all information of
the compositional part \(x_k\) relative to the remaining parts of the
composition of \(\pmb{x}\), termed pivot balance coordinate
\parencite{fivserova2011, hron2012}, is defined as

\begin{equation}
z_k = \sqrt{\frac{D - k}{D - k + 1}} \ln \frac{x_k}{\sqrt[D-k]{\prod_{i=k+1}^{D} x_i}}, \quad k = 1, \ldots, D - 1.
\end{equation}
Table \ref{tab-sbp} gives an example of a complete SBP used to construct pivot coordinates from a five-part composition $\pmb{x}_{ij} = (x_{1ij}, x_{2ij}, x_{3ij}, x_{4ij}, x_{5ij})$.
With this specific choice of coordinates, all relative information about the first part $x_{1ij}$ (pivot element) is contained exclusively in the coordinate $z_{1ij}$, but not in the other coordinates.
If one were interested in an interpretation about another part, 
for example $x_{2ij}$, the role of $x_{1ij}$ and 
$x_{2ij}$ is exchanged by placing $x_{2ij}$ to the first position in the compositional vector, and 
the same type of coordinate is constructed. The resulting coordinates are, thus, rotations of the 
original coordinates. In this way, from a $D$-part composition,
we can construct $D$ pivot coordinates for the
compositional parts of interest, which are all rotations of each other,
and where only the first
coordinate (pivot balance) is used for an interpretation of the respective part.

\begin{table*}[!htbp]
  \caption{Example Sequential Binary Partition of A D = 5 Compositional Parts to Construct (D -1) = 4Pivot Balance Coordinates.}
  \centering
    {\renewcommand{\arraystretch}{1.25}
    \setlength\tabcolsep{0pt}
    \begin{tabular*}{\linewidth}{@{\extracolsep{\fill}} crrrrr}
\toprule
Coordinate & $x_{1ij}$ & $x_{2ij}$ & $x_{3ij}$ & $x_{4ij}$ & $x_{5ij}$ \\
\midrule
1 & +1 & -1 & -1 & -1 & -1 \\
2 &  0 & +1 & -1 & -1 & -1 \\
3 &  0 &  0 & +1 & -1 & -1 \\
4 &  0 &  0 &  0 & +1 & -1 \\
\bottomrule
  \end{tabular*}}
\label{tab-sbp}
\end{table*}

Regardless of the choice of SBP, the ilr
coordinates are linearly independent multivariate real values (if the
compositional parts are strictly positive) and overcome
multicollinearity. Therefore, they can be entered in conventional
statistical models (e.g., multilevel models), making them tractable and
easy to understand. Importantly, the ilr transformation is invertible,
such that the ilr coordinates can be back-transformed via their 1 - 1
relationship to the original composition, as required
\parencite{egozcue2005}.

\subsection{Multilevel Compositional Data and Transformations}
Compositional data (e.g., activity,
diet) may be measured on multiple people \(j = 1,2,\ldots, J\) at
multiple time points \(i = 1,2,\ldots, I\).
We denote such data as
\(\pmb{x}_{ij} = (x_{1ij}, x_{2ij}, \ldots, x_{Dij})\), which is
a vector of compositional data observed at the \(i^{\text{th}}\) time
point for the \(j^{\text{th}}\) person. Therefore,
\(\pmb{x}_{ij}\) can vary between individuals and across time
points within an individual, containing both between-person and
within-person variability \parencite{Curran2011}. We express the \(D\)-part,
time-varying multilevel composition \(\pmb{x}_{ij}\) as

\begin{equation}
\begin{aligned}
\pmb{x}_{ij} 
&= 
\left( x_{1ij}, x_{2ij}, \ldots, x_{Dij} \right)
\\
&= 
\mathcal{C} 
\begin{pmatrix}
x{^{(b)}_{1 \pmb{\cdot} j}} \pmb{\cdot} x{^{(w)}_{1ij}}, 
x{^{(b)}_{2 \pmb{\cdot} j}} \pmb{\cdot} x{^{(w)}_{2ij}}, 
\ldots, 
x{^{(b)}_{D \pmb{\cdot} j}} \pmb{\cdot} x{^{(w)}_{Dij}}  
\end{pmatrix}
\\
&=
\pmb{x}{^{(b)}_{\pmb{\cdot} j}} \oplus \pmb{x}{^{(w)}_{ij}}
\end{aligned}
\end{equation}
where
\begin{itemize}
\item
  \(x{^{(b)}_{d \pmb{\cdot} j}}\) is the person-specific mean of
  the \(d^{\text{th}}\) compositional part over time, which contains
  only between-person variance and no within-person variance. The
  subscript \(\pmb{\cdot}j\) denotes the average across \(i\)
  observations for the individual \(j\) and superscript \((b)\) denotes
  the \(between\) component of the compositional parts.
\item
  \(x{^{(w)}_{dij}}\) is the time-specific deviation of the
  \(d^{\text{th}}\) compositional part from the person \(j\) specific
  mean (i.e., compositional mean-centered deviate), which has
  within-person variance and no between-person variance. The superscript
  \((w)\) denotes the \(within\) component of the composition parts.
\item
  \(\mathcal{C}\) is the closure operation, and
\item
  \(\oplus\) is the perturbation operation on the simplex.
\end{itemize}
The between- and within-person subcompositions can also be expressed as
compositions themselves as

\begin{equation}
\begin{aligned}
\pmb{x}{^{(b)}_{ \pmb{\cdot} j}}    
&= 
\mathcal{C} \left(
x{^{(b)}_{1 \pmb{\cdot} j}},
x{^{(b)}_{2 \pmb{\cdot} j}},
\ldots,
x{^{(b)}_{D \pmb{\cdot} j}} \right) \text{ and } \\
\pmb{x}{^{(w)}_{ij}}          
&= 
\mathcal{C} \left(x{^{(w)}_{1ij}}, x{^{(w)}_{2ij}}, \ldots, x{^{(w)}_{Dij}}\right)
\end{aligned}
\end{equation}

The ilr transformed coordinates
\(\pmb{z}_{ij} \in \mathbb{R}^{D - 1}\) corresponding to the
composition \(\pmb{x}_{ij} \in \mathcal{S}^D\) can also be
uniquely (with respect to the specific ilr transformation) decomposed
into its between- and within-person components in a more familiar
additive way

\begin{equation}
\begin{aligned}
\pmb{z}_{ij} 
& = 
\left(z_{1ij}, z_{2ij}, \ldots, z_{(D - 1)ij} \right)
\\
& =
\left( z{^{(b)}_{1    \pmb{\cdot} j}} + z{^{(w)}_{1ij}}, 
z{^{(b)}_{2     \pmb{\cdot} j}} + z{^{(w)}_{2ij}}, 
\ldots, 
z{^{(b)}_{(D-1) \pmb{\cdot} j}} + z{^{(w)}_{(D-1)ij}} \right)
\\
&=
\pmb{z}{^{(b)}_{\pmb{\cdot} j}} + \pmb{z}{^{(w)}_{ij}}
\end{aligned}
\end{equation}
in which \(z_{kij}\) is the value of the \(k^{\text{th}}\)
(\(k = 1,2,\ldots, D-1\)) ilr coordinate at time point \(i\) for
individual \(j\) and superscripts \((b)\) and \((w)\) denote the between
and within components, respectively, of the ilr coordinates.
Although we focused on longitudinal data (repeated measures are nested
within person) here, the same principles can also be used to distinguish
within- and between-person effects in hierarchical data (individuals are
nested within groups). In any applications, \(i\) index the elementary
``level 1'' units and \(j\) index the clusters or ``level 2'' units. It
should be noted that the separation of within-person and between-person
effects only works to two-level data structure, wherein between-person
level is person-mean at level 2, and within-person level is the
mean-centered deviate at level 1.

Disaggregating effects for more-than-two-level models are not 
currently supported in \textbf{\textit{multilevelcoda}}. Methods research 
has generally not explored how to dissaggregate multilevel models
beyond two levels. For now, we recommend keeping the
data at the aggregate level (that is, not separated by between and
within-person effects), while considering appropriate interpretation
\parencite[see][for a discussion on between-person and within-person inferences]{Curran2011}.

\section{Model Description}

As we adopt Bayesian inference from a pragmatic perspective, our exposition of
it is kept to a minimum. Readers interested in further
methodological guidance on Bayesian analyses are referred to
\cite{kruschke2014, mcelreath2018} for introductions and
\cite{gelman2013bayesian, burkner2017} for more advanced usage. In the
following section, multilevel models with compositional predictors and
their associated post-hoc substitution analyses are first described,
followed by multilevel models with compositional responses.

\subsection{Multilevel Models with Compositional Predictors}

To express a linear model for the time-varying \(D\)-part multilevel
compositional predictor, we first denote the outcome variable observed
at time point \(i\) for individual \(j\) as \(y_{ij}\). The prediction
of a continuous, normally distributed outcome \(y_{ij}\) is the linear
combination of the between-person and within-person effects of a
\(D\)-part composition (expressed as a set of \((D - 1)\)-dimension ilr
coordinates). A linear multilevel model of \(y_{ij}\) can be written as

\begin{equation}
y_{ij} =
  \beta_{0j} + 
  \overbrace{\sum_{k = 1}^{D-1} \beta_k z^{(b)}_{k \pmb{\cdot} j}}^{\text{between}} + 
  \underbrace{\sum_{k = 1}^{D-1}\beta_{(k + D - 1),j}\hspace{0.1cm} z^{(w)}_{kij}}_{\text{within}} +
  \varepsilon_{ij}
\end{equation}
where
\[
\begin{aligned}
\begin{bmatrix}
    \beta_{1} \\
    \vdots\\
    \hspace{0.145cm}\beta_{(D-1)\hspace{0.145cm}}
\end{bmatrix}
&=
\begin{bmatrix}
    \gamma_{1} \\
    \vdots\\
    \hspace{0.08cm}\gamma_{(D-1)\hspace{0.07cm}}
\end{bmatrix} \\
\begin{bmatrix}
    \beta_{0j} \\
    \beta_{Dj}    \\
    \vdots            \\
    \beta_{2(D - 1)j}
\end{bmatrix}
&=
\begin{bmatrix}
    \gamma_{0}   \\
    \gamma_{D}    \\
    \vdots            \\
    \gamma_{2(D - 1)}
\end{bmatrix}
+
\begin{bmatrix} 
    u_{0j} \\
    u_{1j}            \\
    \vdots            \\
    u_{(D - 1)j}
\end{bmatrix} \\
\begin{bmatrix} 
    u_{0j} \\
    u_{1j}            \\
    \vdots            \\
    \hspace{0.13cm}u_{(D - 1)j}\hspace{0.13cm}
  \end{bmatrix}     
& \sim \text{MVNormal}(\pmb{0}, \pmb{\Sigma_u}) \\
  \varepsilon_{ij}       & \sim \text{Normal}(0,      \sigma^{2}_{\varepsilon})
\end{aligned}
\]
The \(\gamma\)s are the population-level effects, \(u\)s are group-level effects, and \(\pmb{\Sigma_u}\) is a variance-covariance matrix for the group-level effects. 
The between- and within-person components of the composition (expressed
as a set of ilr coordinates) are \(z^{(b)}_{\pmb{\cdot} j}\) and
\(z^{(w)}_{ij}\), with the subscripts denoting that the between
component is unique to individual \(j\) and the within component is
unique to time \(i\) for individual \(j\). Thus, all
\(z^{(b)}_{\pmb{\cdot} j}\) and \(z^{(w)}_{ij}\) can be included
as population-level effects (\(\gamma\)), and only
\(z^{(w)}_{ij}\) can be included as group-level effects (
\(u\)). The between- and within-person effects of the \(k^{\text{th}}\)
\((k =1,2,\ldots,D-1)\) ilr coordinates are \(\beta_k\) and
\(\beta_{k + D - 1}\), respectively. Because each ilr coordinate is
decomposed into its between- and within-person components, the total
number of \(\beta\) parameters for the ilr coordinates is twice the
number of the ilr coordinates. 
Further (time varying) population- and/or group-level 
covariates are not included here but can easily be incorporated.

\subsection{Multilevel Compositional Substitution Analysis}

When examining the relationships between compositional predictors and an
outcome, we often are interested in the expected difference in the
outcome when a fixed amount of the composition is reallocated from one
compositional component to another, while the other components remain
constant. These changes can be examined using isotemporal compositional
substitution analysis. In the following, we describe this model in the
multilevel framework.

\subsubsection{Prediction for A Given (Reference) Composition}

For a \(D\)-part composition for person \(j\) at time \(i\),
\(\pmb{x}_{ij} = \pmb{x}_{{\cdot}j}^{(b)} {\oplus} \hspace{0.1cm} \pmb{x}_{ij}^{(w)}\),
and the corresponding set of ilr coordinates
\(\pmb{z}_{ij}=\pmb{z}_{{\cdot}j}^{(b)} + \hspace{0.1cm} \pmb{z}_{ij}^{(w)}\),
the predicted \(y_{ij}\) is

\begin{equation}
  \hat{y}_{ij} =
  \hat{\beta}_{0j} +
  \sum_{k = 1}^{D-1} \hat{\beta}_k               z{^{(b)}_{k \pmb{\cdot} j}} +
  \sum_{k = 1}^{D-1} \hat{\beta}_{(k + D - 1),j} z{^{(w)}_{kij}}
\end{equation}
Now consider the reallocation of a given amount from one part of the
composition, denoted \(d\), to another part, denoted \(d'\), 
where \(d' \neq d \in \{1, \ldots, D\}\).
This is only possible with reference to a
starting composition. The starting composition where compositional
components are reallocated from/to is referred to as the reference
composition (commonly the \textit{compositional mean}, although any
reference composition could be used). The decomposition of a reference
composition \(\pmb{x}_{0}\) is

\begin{equation}
\begin{aligned}
  \pmb{x}_{0}
  &=
  \pmb{x}{^{(b)}_{0}} \oplus \pmb{x}{^{(w)}_{0}} \\
  &=
  \mathcal{C}
  \left(
  {x}^{(b)}_{10}   \pmb{\cdot} {x}^{(w)}_{10},
  \ldots,
  {x}^{(b)}_{d0}   \pmb{\cdot} {x}^{(w)}_{d0},
  \ldots,
  {x}^{(b)}_{d'0}  \pmb{\cdot} {x}^{(w)}_{d'0},
  \ldots,
  {x}^{(b)}_{D0}   \pmb{\cdot} {x}^{(w)}_{D0}
  \right)
\end{aligned}
\end{equation}
Note when the reference composition is a compositional mean value at the
between-person level, the within-person subcomposition
\(\pmb{x}{^{(w)}_{0}}\) becomes the neutral element of the
simplex,
\(\pmb{1}_D = \mathcal{C}(1, 1, \ldots, 1) = (\kappa/D, \kappa/D, \ldots, \kappa/D)\)
as there is no within-person deviation. In such cases, the reference
composition and its corresponding ilr transformation can be simplified
to

\[
\begin{aligned}
\pmb{x}_{0}
&=\pmb{x}^{(b)}_{0} \oplus \pmb{1}_{D}
=\pmb{x}^{(b)}_{0}
\end{aligned}
\]
The predicted outcome at a reference composition \(\pmb{x}_{0}\)
is

\begin{equation}
  \begin{aligned}
    \hat{y}_{0} =
    \hat{\beta}_{0j} +
    \sum_{k = 1}^{D-1} \hat{\beta}_k               z{^{(b)}_{k0}} +
    \sum_{k = 1}^{D-1} \hat{\beta}_{(k + D - 1),j} z{^{(w)}_{k0}}
  \end{aligned}
\end{equation}
where \(z{^{(b)}_{k0}}\) and \(z{^{(w)}_{k0}}\) are the between- and
within-person ilr coordinates at the reference composition,
respectively.

Given that multilevel composition contains both between- and
within-person variability, we can investigate the changes in the outcome
associated with the reallocation of compositional parts at between- and
within-person levels. There are important distinctions between the two
approaches. A between-person substitution examines the differences in
the outcome between individuals with different mean compositions,
whereas a within-person substitution examines the differences in the
outcome associated with the changes in the composition within an
individual (i.e., the deviations from their own mean composition).

\subsubsection{Between-person Substitution}
We denote the two compositional parts involved in a given between-person pairwise substitution as 
$x{^{(b)}_{d0}}$ and $x{^{(b)}_{d'0}}$.
The reallocation of a fixed amount \(t\) from
\(x{^{(b)}_{d0}}\) to \(x{^{(b)}_{d'0}}\) (that is, adding \(t\) to
\(x{^{(b)}_{d'}}\) and subtracting \(t\) from \(x{^{(b)}_{d0}}\)
simultaneously) around a reference composition \(\pmb{x}_{0}\) at
the between-person level is

\begin{equation}
  \begin{aligned}
    x{^{{(b)}'}_{d}}  &= x{^{(b)}_{d0}}  - t \\
    x{^{{(b)}'}_{d'}} &= x{^{(b)}_{d'0}} + t
  \end{aligned}
\end{equation}
where \(d' \neq d \in \{1, \ldots, D\}\), \(t\) is the reallocated
change (e.g., minutes/1440 if \(\kappa=1440\)), and
\(0 < t < \min \left\{ x^{(b)}_{d}, \kappa - x^{(b)}_{d'} \right\}\).
Keeping the remaining parts of the composition constant, the new
\(D\)-part composition \(\pmb{x}{^{{(b)}'}_{(d-d')}}\) can be expressed
as

\begin{equation}
  \begin{aligned}
    \pmb{x}{^{{(b)}'}_{(d-d')}}
      & =
    \mathcal{C}(
    x{^{(b)}_{10}}           \pmb{\cdot}   x{^{(w)}_{10}},
    \ldots,
    x{^{{(b)}'}_{d}}         \pmb{\cdot}   x{^{(w)}_{d0}},
    \ldots,
    x{^{{(b)}'}_{d'}}        \pmb{\cdot}   x{^{(w)}_{d'0}},
    \ldots,
    x{^{(b)}_{D0}}           \pmb{\cdot}   x{^{(w)}_{D0}}) \\
      & =
    \mathcal{C}(
    x{^{(b)}_{10}}           \pmb{\cdot}   x{^{(w)}_{10}},
    \ldots,
    (x{^{(b)}_{d0}}  - t)    \pmb{\cdot}   x{^{(w)}_{d0}},
    \ldots,
    (x{^{(b)}_{d'0}} + t)    \pmb{\cdot}   x{^{(w)}_{d'0}},
    \ldots,
    x{^{(b)}_{D0}}           \pmb{\cdot}   x{^{(w)}_{D0}})
  \end{aligned}
\end{equation}
The predicted outcome at the between-person reallocation is given as

\begin{equation}
  \begin{aligned}
    \hat{y}^{{(b)}'}_{(d-d')}
    =
    \hat{\beta}_{0j} +
    \sum_{k = 1}^{D-1} \hat{\beta}_k               z{^{{(b)}'}_{k0}} +
    \sum_{k = 1}^{D-1} \hat{\beta}_{(k + D - 1),j} z{^{(w)}_{k0}}
  \end{aligned}
\end{equation}
where \(z{^{{(b)}'}_{k0}}\) indicates the new between-person ilr
coordinates resulted from the between-person reallocation in the
composition
\(z{^{(w)}_{k0}}\) (within-person ilr coordinates) is the same as the reference ilr coordinates.
The predicted difference in the outcome, \(\Delta{\hat{y}^{(b)}}_{(d-d')}\), for
the between-person changes in compositional parts (i.e., between the
reference composition and the reallocated composition at between-person
level) is therefore

\begin{equation}
  \begin{aligned}
    \Delta{\hat{y}^{(b)}}_{(d-d')}
      & = \hspace{.5cm}
    \hat{y}^{{(b)}'}_{(d-d')}
    -
    \hat{y}_{0} \\
      & = \hspace{.5cm}
    \left(
    \hat{\beta}_{0j} +
    \sum_{k = 1}^{D-1} \hat{\beta}_k               z{^{{(b)}'}_{k0}} +
    \sum_{k = 1}^{D-1} \hat{\beta}_{(k + D - 1),j} z{^{(w)}_{k0}}
    \right) \\
      & \hspace{.5cm}
    -
    \left(
    \hat{\beta}_{0j} +
    \sum_{k = 1}^{D-1} \hat{\beta}_k               z{^{(b)}_{k0}} +
    \sum_{k = 1}^{D-1} \hat{\beta}_{(k + D - 1),j} z{^{(w)}_{k0}}
    \right) \\
      & = \hspace{.5cm}
    \sum_{k = 1}^{D-1} \hat{\beta}_k \left(z{^{{(b)}'}_{k0}} - z{^{(b)}_{k0}}\right) \\
  \end{aligned}
\end{equation}

\subsubsection{Within-person Substitution}

The reallocation of a fixed amount \(t\) between two compositional parts
at the within-person level (\(x{^{(w)}_{d0}}\) and \(x{^{(w)}_{d'0}}\))
around a reference composition can be expressed as

\begin{equation}
  \begin{aligned}
    x{^{{(w)}'}_{d}}  & = x{^{(w)}_{d0}}  - t \\
    x{^{{(w)}'}_{d'}} & = x{^{(w)}_{d'0}} + t .
  \end{aligned}
\end{equation}
The new \(D\)-part composition for within-person level reallocation of
\(t\) becomes

\begin{equation}
  \begin{aligned}
    \pmb{x}{^{{(w)}'}_{(d-d')}} 
      & =
    \mathcal{C}(
    x{^{(b)}_{10}}          \pmb{\cdot}  x{^{(w)}_{10}},
    \ldots,
    x{^{(b)}_{d0}}          \pmb{\cdot}  x{^{{(w)}'}_{d}},
    \ldots,
    x{^{(b)}_{d'0}}         \pmb{\cdot}  x{^{{(w)}'}_{d'}},
    \ldots,
    x{^{(b)}_{D0}}          \pmb{\cdot}  x{^{(w)}_{D0}}) \\
      & =
    \mathcal{C}(
    x{^{(b)}_{10}}          \pmb{\cdot}  x{^{(w)}_{10}},
    \ldots,
    x{^{(b)}_{d0}}          \pmb{\cdot}  (x{^{(w)}_{d0}}  - t),
    \ldots,
    x{^{(b)}_{d'0}}         \pmb{\cdot}  (x{^{(w)}_{d'}} + t),
    \ldots,
    x{^{(b)}_{D0}}          \pmb{\cdot}  x{^{(w)}_{D0}}).
  \end{aligned}
\end{equation}
The predicted outcome for the within-person reallocation is
\begin{equation}
  \begin{aligned}
    \hat{y}^{(w)'}_{(d-d')}
    =
    \hat{\beta}_{0j} +
    \sum_{k = 1}^{D-1} \hat{\beta}_k               z{^{(b)}_{k0}} +
    \sum_{k = 1}^{D-1} \hat{\beta}_{(k + D - 1),j} z{^{{(w)}'}_{k0}}
  \end{aligned}
\end{equation}
where the \(z{^{(b)}_{k0}}\) remains the same as the reference
between-person ilr coordinates, whereas the \(z{^{{(w)}'}_{k0}}\) is the
new within-person ilr coordinates, showing the change in within-person
ilr coordinates relative to the reference point.
Thus, the predicted changes in the outcome due to the changes across the
compositional parts at the within-person level,
\(\Delta{\hat{y}^{(w)}_{(d-d')}}\), is

\begin{equation}
  \begin{aligned}
    \Delta{\hat{y}^{(w)}_{(d-d')}} 
      & = \hspace{.5cm}
    \hat{y}^{(w)'}_{(d-d')}
    -
    \hat{y}_{0} \\
      & =
    \hspace{.5cm}
    \left(
    \hat{\beta}_{0j} +
    \sum_{k = 1}^{D-1} \hat{\beta}_k                z{^{(b)}_{k0}} +
    \sum_{k = 1}^{D-1} \hat{\beta}_{(k + D - 1),j}  z{^{{(w)}'}_{k0}}
    \right) \\
      & \hspace{.5cm}
    -
    \left(
    \hat{\beta}_{0j} +
    \sum_{k = 1}^{D-1} \hat{\beta}_k                z{^{(b)}_{k0}} +
    \sum_{k = 1}^{D-1} \hat{\beta}_{(k + D - 1),j}  z{^{(w)}_{k0}}
    \right) \\
      & = \hspace{.5cm}
    \sum_{k = 1}^{D-1} \hat{\beta}_{(k + D - 1),j}
    \left (z{^{{(w)}'}_{k0}} - z{^{(w)}_{k0}} \right).
  \end{aligned}
\end{equation}

\subsubsection{Substitution Analysis Framework}
We propose two frameworks for the substitution analysis (Table
\ref{tab-sub}), with noteworthy distinctions. The \textbf{Simple substitution
analysis} provides simple effects of the change in a composition on
an outcome, where the reference composition could be grand compositional mean 
or any (constant) hypothetical set of values.
The \textit{Average substitution analysis} is motivated by average marginal effects
\parencite{norton2019, mize2019}. That is, using the cluster (e.g., person)
compositional mean as the reference composition to estimate the
predicted changes in the outcome for each cluster, then averaging across
the prediction to obtain the average change of the sample. This estimate
reflects the change in outcome when every cluster (e.g., person) in the
sample reallocates a \(t\) unit from one compositional part to another,
which demonstrates the change for the full distribution of the
predictor(s) rather than an arbitrary prediction \parencite{leeper2017}.

For linear outcomes, the results produced by the two models are expected
to be comparable. \textit{Average substitution analysis} provides better
estimates than \textit{Simple substitution analysis} particularly in the cases of
non-linear outcomes, models with covariates, large reallocation across
compositional parts resulting in one part approaching zero, or
imbalanced data, such as the unequal balance of time spent in sleep-wake
behaviours across individuals (e.g., shift workers vs non-shift
workers, male vs females). In addition, the credible intervals estimated
by the \textit{Simple substitution analysis} only reflect the population level
effects, whereas the credible intervals estimated by the average
substitution analysis incorporate the variability at the group-level by
including all group-level effects.

\textit{Average substitution analysis} generally require more time and
computational resources than \textit{Simple substitution analysis}, as the
estimation takes place at the cluster-level. However, all
\texttt{substitution()} analyses can be executed in parallel using available
\textbf{R} packages, such as \textbf{\textit{doFuture}}
\parencite{doFuture}, to optimise computational time and performance.

\begin{table*}[!htbp]
\caption{Substitution analysis framework.}
    \centering
    \begin{tabular*}{\textwidth}{m{1\textwidth}}
    \textit{Simple substitution analysis} \\
    \toprule
      Examines the change in the outcome for the \textit{between-person} and 
      \textit{within-person} reallocation of compositional parts, 
      using the \textbf{grand compositional mean} or a \textbf{user's specified composition} as the reference composition. 
      Estimated change in outcome is the simple effect of the compositional reallocation, 
      and if incorporated, at different levels of the other (categorical) predictors or is an unweighted average of them. \\
      The fitting procedure of the \textit{Simple substitution analysis} consists of \textbf{6} main steps:
      \begin{enumerate}
      \item Calculate the reference composition ($\pmb{x}_0$) (e.g., grand compositional mean) and 
            its corresponding ilr coordinates ($\pmb{z}_0$)
      \item Estimate the outcome at the reference composition, $\hat{y}_0$.
      \item Using the reference composition, generate new composition(s),$\pmb{x}{'_{0}}$, 
            and the corresponding ilr coordinates, $\pmb{z}{'_{0}}$, 
            for the reallocation(s) of $t$ from one part of the composition to another.
      \item Estimate the outcome at the reallocated compositions, $\hat{y}'_0$.
      \item Calculate the changes in the outcome for the $t$ reallocation(s), ($\Delta{\hat{y}}$).
      \item Repeat this procedure for all compositional parts (end after $D$ steps).
      \end{enumerate} \\
      \textit{Average substitution analysis}  \\
   \midrule
      Examines the average change in the outcome for the \textit{between-person} and \textit{within-person} 
      reallocation of compositional parts, using the 
      \textbf{cluster (e.g., individual) compositional mean} as the reference composition. 
      Change in the outcome is estimated for each cluster then averaged over the data to obtain an average change of the compositional
      reallocation, which gives weighted prediction over the empirical (sample) distribution. \\
      The estimation of \textit{Average substitution analysis} follows \textbf{7} main steps:
      \begin{enumerate}
      \item Calculate the reference composition ($\pmb{x}_0$) (i.e., cluster compositional mean) 
            and its corresponding ilr coordinates ($\pmb{z}_0$).
      \item Estimate the outcome at the cluster compositional mean, $\hat{y}_0$.
      \item Using the cluster compositional mean, generate new composition(s)
            $\pmb{x}{'_{0}}$,
            and the corresponding ilr coordinates, $\pmb{z}{'_{0}}$,
            for the reallocation(s) of $t$ from one part of the composition to another for each cluster.
      \item Estimate the outcome at the reallocated compositions for each cluster, $\hat{y}'_0$.
      \item Calculate the changes in the outcome for the $t$ reallocation(s), 
            ($\Delta{\hat{y}}$), at the cluster level.
      \item Average the changes in $\Delta{\hat{y}}$, that is $\overline{\Delta{\hat{y}}}$ over the clusters/data.
      \item Repeat this procedure for all compositional parts (end after $D$ steps).
      \end{enumerate} \\
    \bottomrule
  \end{tabular*} \\
  \label{tab-sub}
\end{table*}

\section{Package Overview}
Package \textbf{\textit{multilevelcoda}} provides functions for fitting multivariate
multilevel models with compositional data using full Bayesian inference.
The package is open-source software for the \textbf{R} programming
platform. The latest release version of \textbf{\textit{multilevelcoda}} from the
Comprehensive \textbf{R} Archive Network (CRAN) can be installed via
\texttt{install.packages("multilevelcoda")}. Alternatively, the current
developmental version can be downloaded from GitHub via

\begin{lstlisting}
R> devtools::install_github(
+ "florale/multilevelcoda")
\end{lstlisting}

\textbf{\textit{multilevelcoda}} uses the \textbf{R} package \textbf{\textit{brms}} to build
models, which in turn uses the probabilistic programming language
\textbf{Stan} as the backend that dynamically generates and compiles
\textbf{C++} code for specific, Bayesian models. Thus, a
\textbf{C++} compiler is required, beyond just having a functional
\textbf{R} installation. For Windows, the program \textbf{\textit{Rtools}}
\parencite{rtools} comes with a \textbf{C++} compiler. On Mac,
Installation of \textbf{Xcode} \parencite{xcode} for Mac, is required. Linux
requires \textbf{g++} or \textbf{Clang}. Detailed instructions on how to get
the compilers and running can be found in the prerequisites section on
the
\href{https://github.com/stan-dev/rstan/wiki/RStan-Getting-Started}{RStan
package's website}. Note that the \textbf{\textit{rstan}} package
\parencite[the \textbf{R} interface of \textbf{Stan};][]{rstan} also
depends heavily on several other R packages; these dependencies are
automatically installed if the \textbf{\textit{rstan}} package (R interface to
Stan) is installed via one of the conventional mechanisms.
Users will find further assistance through \textbf{R} documentation
and vignettes to guide them through the functionality of the package, as
well as examples of its use. Contributions are welcome both in terms of
bug reports and feature enhancements, via the standard mechanism of
GitHub issues and pull requests.

\begin{figure*}[!htbp]
\caption{Implementing the Bayesian multilevel compositional data analysis using R package \textbf{\textit{multilevelcoda}}.}
\centering
\tikzstyle{process} = [rectangle, rounded corners, text width=5cm, minimum height=1.5cm, text centered, draw=black, thick, fill=lightblue, font=\small]
\tikzstyle{fun}     = [rectangle, rounded corners, text width=3cm, minimum height=1.5cm, text centered, draw=black, thick, fill=midblue, text=black, font=\small]
\tikzstyle{line} = [thick,-,>=stealth]
\tikzstyle{arrow} = [thick,->,>=stealth]

\begin{tikzpicture}[node distance = 2.25cm]

\node (pro1) [process, align=center]                {Compute multilevel compositional data and log-ratio transforms};
\node (pro2) [process, align=center, below of=pro1] {Fit Bayesian (multivariate) multilevel models for compositional data};
\node (pro3) [process, align=center, below of=pro2] {Estimate pivot coordinates};
\node (pro4) [process, align=center, below of=pro3] {Run multilevel compositional substitution analysis};
\node (pro5) [process, align=center, below of=pro4] {Present and interpret final results};

\node (pro6) [fun, align=center,                xshift = 5 cm] {\texttt{complr()}};
\node (pro7) [fun, align=center, below of=pro1, xshift = 5 cm] {\texttt{brmcoda()}};
\node (pro8) [fun, align=center, below of=pro2, xshift = 5 cm] {\texttt{pivot\_coord()}};
\node (pro9) [fun, align=center, below of=pro3, xshift = 5 cm] {\texttt{substitution()}};
\node (pro10)[fun, align=center, below of=pro4, xshift = 5 cm] {\texttt{summary()} and \texttt{plot()}};

\draw [arrow] (pro1) -- (pro2);
\draw [arrow] (pro2) -- (pro3);
\draw [arrow] (pro3) -- (pro4);
\draw [arrow] (pro4) -- (pro5);

\draw [line] (pro1) -- (pro6);
\draw [line] (pro2) -- (pro7);
\draw [line] (pro3) -- (pro8);
\draw [line] (pro4) -- (pro9);
\draw [line] (pro5) -- (pro10);

\end{tikzpicture}
\label{fig-workflow}
\end{figure*}

Analysis in \textbf{\textit{multilevelcoda}} follows the procedure in
Figure \ref{fig-workflow}. First, the user calculates the composition
and the corresponding log-ratio transforms (e.g., ilr coordinates) using
\texttt{complr()} function. Next, this information is used to fit Bayesian
(multivariate) multilevel models using the \texttt{brmcoda()} function. During
this step, the model is passed to \texttt{brm()} to generate
\textbf{Stan} model code, which is then passed to either the
\textbf{\textit{rstan}} package \parencite{rstan} or the \textbf{\textit{cmdstanr}} package
\parencite[the \textbf{R} interface of \textbf{\textit{CmdStan}};][]{cmdstanr}. Models
are compiled in \textbf{C++}, fitted by \textbf{Stan}, and
post-processed in \textbf{\textit{brms}} before being saved in
\textbf{\textit{multilevelcoda}}'s \texttt{brmcoda()} in \textbf{R}. The results
from \texttt{brmcoda} can be used to estimate the
pivot coordinates of the composition using the \texttt{pivot\_coord()} function, and
substitution analysis using the \texttt{substitution()} function. Finally,
results from all functions can
be investigated in \textbf{R} using various methods such as
\texttt{summary()}, \texttt{plot()}, or \texttt{predict()} (for example, for a
complete list of methods defined on the \texttt{brmcoda} object, type
\texttt{methods(class = "brmcoda")}).

\section{Example Application}
This section presents examples to implement 
Bayesian multilevel compositional data analysis following workflow in Figure \ref{fig-workflow}.
Reproducible code can be found at
\url{https://github.com/florale/multilevelcoda-overview}.
The example data set is a simulated, built-in data set in a long format, with repeated measurements of stress and
24h time use separated into five behaviours: total sleep time, time awake in bed, moderate-to-vigorous physical activity (MVPA), light physical activity (LPA), and sedentary behaviour (SB). Daily stress was measured on a 0-10 scale. The five behaviours make up a 5-part composition.
\begin{lstlisting}
R> library(multilevelcoda)
R> data(mcompd)
R> data(psub)
\end{lstlisting}
The example data set \texttt{mcompd} is in long format and consists of
3540 entries of 10 variables

\begin{lstlisting}
R> head(mcompd)
ID Stress  TST WAKE MVPA  LPA   SB Time Age Female
185     4  542   99  297  460   41    1  30      0
185     7  458   49  117  653  162    2  30      0
185     3  271   41  489  625   15    3  30      0
185     2  525   76  259  398  182    4  30      0
185     8  651   86  112  436  155    5  30      0
185     8  431   84  264  476  185    6  30      0
\end{lstlisting}

Variable \texttt{ID} is participant id. \texttt{Stres} is self-reported
stress measured on a 0-10 scale. Sleep duration (i.e., total sleep time,
\texttt{TST}), time awake in bed (\texttt{WAKE}), moderate-to-vigorous
physical activity (\texttt{MVPA}), light physical activity (\texttt{LPA}),
and sedentary behaviour (\texttt{SB}) make up a 5-part composition.
\texttt{Time} is time point id at which stress and the 5-part composition
were repeatedly measured. Finally, variables \texttt{Age} and
\texttt{Female} are individual baseline factors.

\subsection{Transforming Multilevel Compositional Data}
The \texttt{complr()} function processes compositional data and performs
log-ratio transformation. 
First, to build a set of ilr coordinates, a SBP is required.
The construction and interpretation of $ilr$ coordinates may depend on 
specific application.
Alternatively, multilevelcoda uses pivot coordinates as the default SBP,
where the first pivot coordinate represent the ratio of
the first compositional part relative to the remaining parts.
We may use the following code to process and transform compositional
data:

\begin{lstlisting}
R> cilr <- complr(
+  data  = mcompd,
+  parts = c("TST", "WAKE", "MVPA", "LPA", "SB"),
+  idvar = "ID",
+  total = 1440
+ )
\end{lstlisting}
We specify the variable that identifies how units are clustered 
\texttt{idvar} as \texttt{ID} based on our data, and specify
value of \texttt{total} to be \texttt{1440}, which is the total minutes of a
24-hour day, to which the 5 behaviours must sum. 
In this example (when a SBP is not specified), the default SBP is
\begin{lstlisting}
R> sbp
     TST WAKE MVPA LPA SB
[1,]   1   -1   -1  -1 -1
[2,]   0    1   -1  -1 -1
[3,]   0    0    1  -1 -1
[4,]   0    0    0   1 -1
\end{lstlisting}
where each column represents one of the 5 parts of the composition and each row represents one of the transformed 4 (5-1) ilr coordinates. 
Here, the first coordinate represent the ratio of sleep relative to the remaining behaviours. Intepretation of the coordinates based on this SBP is in Table \ref{tab-pivot-coord}.
A summary the transformed data can be obtained by running

\begin{lstlisting}
R> summary(cilr)                                    
composition_parts  TST, WAKE, MVPA, LPA, SB
logratios            ilr1, ilr2, ilr3, ilr4
idvar                                  NULL
nobs                                   3540
ngrps                                   266
transform_type                          ilr
total                                  1440
composition_geometry                  acomp
logratio_class                        rmult
\end{lstlisting}
The output provides information about the data and transformation. Some
general information on the data include the names of compositional parts
(\texttt{composition\_parts}) and log-ratio variables
(\texttt{logratios}), the ID variable (\texttt{idvar}, for multilevel
dataset), number of observations (\texttt{nobs}) and number of groups
(\texttt{ngrps}). Other information to perform transformation on
compositional data includes transformation methods (either \texttt{ilr},
\texttt{alr}, or \texttt{clr}), the closure value (for ilr
transformation), the geometries and classes of the composition and the
ilr coordinates. The class \texttt{acomp} indicates the composition class
that aligns with the philosophical framework of the Aitchison Simplex,
whereas ilr are real multivariate vectors. Within a \texttt{complr} object
(not shown), data sets of composition and ilr coordinates are stored
alongside the original data set, which are used for subsequent analyses.

\subsection{Fitting Bayesian Multilevel Models with Compositional Predictors}
Multilevel models for compositional data are estimated using
\texttt{brmcoda()}. To examine how between-person and within-person 24h
behaviours predict stress, we may fit the following model:

\begin{lstlisting}
R> m <- brmcoda(
+   complr  = cilr,
+   formula = Stress ~ 
+             bilr1 + bilr2 + bilr3 + bilr4 +
+             wilr1 + wilr2 + wilr3 + wilr4 + 
+             (1 | ID),
+   warmup = 1000, iter = 2000, seed = 123,
+   chains = 4, cores = 4, backend = "cmdstanr"
+ )
\end{lstlisting}
The structure of \texttt{brmcoda()} has two core arguments: the
\texttt{complr} object, which replaces the standard \texttt{data},
argument for models, and a model \texttt{formula}. The \texttt{formula}
argument takes information on the outcomes and predictors of the model,
separated by the \(\sim\). Models are fitted using \textbf{\textit{brms}},
therefore, the syntax follows the form of a \texttt{brmsfit} object and is
similar to \textbf{\textit{lme4}}'s. The left side of the \texttt{formula} is the
outcome, \texttt{Stress} in this example. The right side of the
\texttt{formula} specifies the predictors, including both population-level
and group-level terms, separated by the \(+\). In the present example,
the population-level terms are 4 \texttt{bilr}s representing the set of
ilr coordinates at between-person level and 4 \texttt{wilr}s representing
the ilr coordinates at within-person level. The group-level terms follow
the form \texttt{(coef | group)}, allowing the intercept to vary by
\texttt{ID}. Note that a \texttt{data} argument is not required, as the data
set is supplied by the \texttt{complr} object. Additional arguments used
for \texttt{brm} model function are specified in the \texttt{...} argument,
such as prior specifications and distribution of the response variable. 
If not otherwise specified, default \texttt{brm}
priors and link functions are applied. This example model \texttt{m} is
fitted using 4 chains, each with 2000 iterations including 1000-warmup
iterations for the sampler, running on 4 cores. The model produces 4000
posterior draws using the HMC sampler.
Weakly-informative priors were used 
(see Supplementary materials for prior information),
which play a minimal role in the computation of the posterior distribution,
and maximise the influence of the data.
Student's t distribution was used for 
the fixed intercept, and 
flat priors (improper priors over the reals) were used
for the fixed parameters of the predictors.
The standard deviation parameters of the random intercept and residual were
specified using student's t distributions.

\subsubsection{Model Summary}
The output of \texttt{brmcoda()} is a \textbf{R} \texttt{brmcoda} object
with 2 elements: an fitted \texttt{brm()} model with class \texttt{brmsfit}
and the input data from \texttt{complr()}. A model summary is 
available via
\begin{lstlisting}
R> summary(m)
 Family: gaussian 
  Links: mu = identity; sigma = identity 
Formula: Stress ~ bilr1 + bilr2 + bilr3 + bilr4 + 
         wilr1 + wilr2 + wilr3 + wilr4 + (1 | ID) 
   Data: tmp (Number of observations: 3540) 
  Draws: 4 chains, each with iter = 2000; warmup = 1000; thin = 1;
         total post-warmup draws = 4000

Multilevel Hyperparameters:
~ID (Number of levels: 266) 
              Estimate Est.Error l-95% CI u-95% CI Rhat Bulk_ESS Tail_ESS
sd(Intercept)     1.00      0.06     0.88     1.13 1.00     1573     2899

Regression Coefficients:
          Estimate Est.Error l-95% CI u-95% CI Rhat Bulk_ESS Tail_ESS
Intercept     2.59      0.48     1.59     3.51 1.00     1463     2242
bilr1         0.39      0.43    -0.43     1.26 1.00     1530     2041
bilr2        -0.10      0.17    -0.46     0.23 1.00     1344     1829
bilr3         0.11      0.21    -0.31     0.54 1.00     1469     2333
bilr4        -0.01      0.28    -0.56     0.55 1.00     1463     1907
wilr1        -0.16      0.16    -0.48     0.15 1.00     4646     2869
wilr2        -0.30      0.08    -0.47    -0.14 1.00     6135     3094
wilr3        -0.10      0.08    -0.25     0.06 1.00     3741     2891
wilr4         0.24      0.10     0.04     0.43 1.00     3951     3377

Further Distributional Parameters:
      Estimate Est.Error l-95% CI u-95% CI Rhat Bulk_ESS Tail_ESS
sigma     2.38      0.03     2.33     2.44 1.00     6030     2794

Draws were sampled using sample(hmc). For each parameter, Bulk_ESS
and Tail_ESS are effective sample size measures, and Rhat is the potential
scale reduction factor on split chains (at convergence, Rhat = 1).
\end{lstlisting}

The model output follows the standard output from \texttt{brm()}. The top
of the output shows the general information of the model, followed by
the group-level (random) effects and population-level (fixed) effects.
At the bottom of the output, family specific parameters and
autocorrelation (if incorporated) are also provided. Every parameter is
summarised using the mean (\texttt{Estimate}), standard deviation
(\texttt{Est.Error}) of the posterior distribution (the standard error of
the estimate), and two-sided 95\% Credible intervals based on the
quantiles (\texttt{l-95\% CI} and \texttt{u-95\% CI}) \parencite{burkner2017}. Additional
information about the model were also provided, including \texttt{Rhat}
for information on the convergence of the algorithm and \texttt{Bulk\_ESS}
and \texttt{Tail\_ESS} for effective sample size (ESS).

Model convergence can be evaluated using diagnostic statistic $\hat{R} < 1.05$
\parencite{vehtari2021} 
and ESS $>$ 400 \parencite{vehtari2021}.
Examining the population-level effects of the ilr coordinates, only
\texttt{wilr2} and \texttt{wilr4} have the two-sided 95\% Credible Intervals
not containing zero. Therefore, we have evidence for the association
between \texttt{wilr2} and \texttt{wilr4} and \texttt{Stress}, repsectively.
Recall that the interpretation of the ilr depends on the SBP. For
example, the significant coefficient for \texttt{wilr2} shows that the
increase time awake in bed while proportionally decreasing waking behaviours 
(MVPA, LPA, and SB) on a given day, predicted lower stress (-0.30 {[}95\%CI -0.47, -0.14{]}).

\subsubsection{Estimating and Interpreting Pivot Coordinate Coefficients}
Pivot coordinates represent the relative importance of
each behaviour in the 24h composition 
(with respect to the geometric average of the remaining behaviours).
Pivot coordinates may be an easier interpretation as they are 
always contrasting one part of the composition to all remaining parts.
Using them also makes the results independent of any one SBP specified,
because pivot coordinates can be calculated as a rotation of any given SBP.
Pivot coordinates can be obtained by running

\begin{lstlisting}
R> m_coordinates <- pivot_coord(m, method = "rotate")
+  summary(m_coordinates)
\end{lstlisting}
We supplied the \texttt{pivot\_coord()} function with a 
\texttt{brmcoda} object, and specified \texttt{method = "rotate"} to indicate 
we want to rotate all possible ilr basis matrix to estimate the pivot coordinates 
representing each 24h behaviour.
The results showing the association between the pivot coordinates representing the 
24h behaviours and stress are in Table \ref{tab-pivot-coord}.
Results showed that higher time spent in awake in bed relative to the remaining behaviours 
was associated with -0.25 {[}95\%CI-0.42, -0.08{]} lower stress, whereas 
higher time spent in LPA relative to the remaining behaviours predicted 0.37
{[}0.12, 0.63{]}) higher stress.

\begin{table*}[!htbp]
\caption{Bayesian Multilevel Model with Compositional Predictor Examining 
         the Associations of the 24-hour Sleep-Wake Behaviours and Stress.}
  \centering
  {\renewcommand{\arraystretch}{1}
  \begin{tabular*}{\linewidth}{@{\extracolsep{\fill}}lc@{\extracolsep{\fill}}}
    \toprule
    Pivot Coordinate & \makecell{Posterior mean \\and 95\% credible intervals}\\
    \midrule
    \textbf{Between-person level} &\\
    Sleep vs remaining        & $\begin{matrix}  0.39 \\ [-0.43, 1.26] \end{matrix}$ \\
    Awake in bed vs remaining & $\begin{matrix} -0.20 \\ [-0.57, 0.16] \end{matrix}$ \\
    MVPA vs remaining         & $\begin{matrix}  0.04 \\ [-0.34, 0.41] \end{matrix}$ \\
    LPA vs remaining          & $\begin{matrix} -0.13 \\ [-0.81, 0.58] \end{matrix}$ \\
    SB vs remaining           & $\begin{matrix} -0.11 \\ [-0.52, 0.30] \end{matrix}$ \\
     \midrule
     \textbf{Within-person level} &\\
     Sleep vs remaining       & $\begin{matrix} -0.16      \\ [-0.48,  0.15] \end{matrix}$ \\
     Awake in bed vs remaining       & $\begin{matrix} -0.25^\ast \\ [-0.42, -0.08] \end{matrix}$ \\
     MVPA vs remaining        & $\begin{matrix}  0.05      \\ [-0.09,  0.19] \end{matrix}$ \\
     LPA vs remaining         & $\begin{matrix}  0.37^\ast \\ [ 0.12,  0.63] \end{matrix}$ \\
     SB vs remaining          & $\begin{matrix} -0.01      \\ [-0.15,  0.14] \end{matrix}$ \\
     
     \bottomrule
    \end{tabular*}} \\
  \raggedright{\textit{Notes.} MVPA = moderate-to-vigorous physical activity, LPA = light physical activity, SB = sedentary behaviour. $^\ast$95\% credible intervals not containing 0.}
\label{tab-pivot-coord}
\end{table*}

\subsection{Running Multilevel Compositional Substitution Analysis}
Beyond understanding the independent and compositional association between behaviours and stress,
the changes in stress for different pairwise
reallocation of behaviours 
(e.g., reallocation between MVPA and SB while keeping the remaining fixed) 
can be estimated using
compositional substitution analysis.
The below example shows how to conduct a \textit{Simple substitution analysis} 
(by automating the steps described in Table \ref{tab-sub}) 
to examine the changes in stress associated with behaviour reallocation for 1 to 10 minutes, at
between-person and within-person levels using the using the \texttt{substitution()} function. 
We use the below code

\begin{lstlisting}
R> sub_simple <- substitution(
+   object = m,
+   delta  = 1:10,
+   ref    = "grandmean",
+   level  = c("between", "within")
+ )
\end{lstlisting}
\texttt{substitution()} requires a \texttt{brmcoda} object.
\texttt{delta = 1:10} indicates the estimation of the changes in the outcome
\texttt{Stress} for the reallocation from 1 to 10 minutes across behaviours.
We also specify \texttt{ref = "grandmean"} to indicate \textit{simple substitution analysis}. 
If desired, \texttt{ref} can also take a
reference grid that contains the combination of predictors (i.e.,
reference composition and other covariates) over which predictions are
made. If an user's specified reference grid is not supplied, the default
reference grid \parencite[imported from \texttt{emmeans} package;][]{emmeans}
consisting of average value of numeric predictors and the levels of the
categorical predictors is used. 
The default 95\% credible interval are used here, however, can be any desired intervals, such as \texttt{ci = 0.99}.
As we are interested in both between- and within-person changes, 
we specified \texttt{level = c("between", "within")}.

Results of the Bayesian compositional substitution analyses are in Table \ref{tab-sub}.
At between-person level, none of the results are significant, as the
95\% credible intervals contain 0, 
showing that reallocation between behaviours was not associated
in changes in stress. At within-person level, there were significant
results for the substitution of time awake in
bed and total sleep time (TST), respectively, with other behaviours.
More 10 minutes in time awake in bed at the expense of 
any other behaviours predicted lower stress 
(estimates range from -0.03 {[}95\%CI -0.06, -0.00{]} to -0.04 {[}95\%CI -0.06, -0.02{]}.
The opposite reallocations were also supported,
with reallocation of 10 minutes from time awake in bed to other behaviours 
was associated with higher stress
(estimates range from 0.04 {[}95\%CI 0.01, 0.07{]} to 0.05 {[}95\%CI 0.02, 0.07{]}.
Additionally, less time in TST
predicted -0.03 lower stress {[}95\% CI -0.06, -0.00{]} when compensated by time awake in bed,
but 0.01 higher stress {[}95\% CI 0.00, 0.02{]} when compensated by LPA.

\begin{table*}[!htbp]
\caption{Bayesian Multilevel Compositional Substitution Analysis Estimating the Difference in Stress Associated with Reallocation of 30 minutes across 24-hour Sleep-Wake Behaviours.}
 \centering
  {\renewcommand{\arraystretch}{1}
  \setlength\tabcolsep{0pt}
  \begin{tabular*}{\textwidth}{@{\extracolsep{\fill}} lccccc }
  \toprule
                & $\downarrow \text{Sleep}$
                & $\downarrow \text{Awake in bed}$
                & $\downarrow \text{MVPA}$
                & $\downarrow \text{LPA}$
                & $\downarrow \text{SB}$ \\
      \midrule
      \textbf{Between-person level} &&&&& \\
      $\uparrow \text{Sleep}$
                & -
                & $\begin{matrix}  0.04 \\ [-0.03, 0.11] \end{matrix}$
                & $\begin{matrix}  0.01	\\ [-0.03, 0.04] \end{matrix}$
                & $\begin{matrix}  0.01	\\ [-0.01, 0.03] \end{matrix}$
                & $\begin{matrix}  0.01 \\ [-0.02, 0.04] \end{matrix}$                 \\

      $\uparrow \text{Awake in bed}$
                & $\begin{matrix} -0.03	\\ [-0.10, 0.02] \end{matrix}$
                & -
                & $\begin{matrix} -0.03 \\ [-0.08, 0.03] \end{matrix}$
                & $\begin{matrix} -0.02 \\ [-0.08, 0.03] \end{matrix}$
                & $\begin{matrix} -0.02 \\ [-0.07, 0.02] \end{matrix}$                 \\

      $\uparrow \text{MVPA}$
                & $\begin{matrix} -0.01 \\ [-0.04, 0.03] \end{matrix}$
                & $\begin{matrix}  0.03 \\ [-0.02, 0.09] \end{matrix}$
                & -
                & $\begin{matrix}  0.00	\\ [-0.02, 0.03] \end{matrix}$
                & $\begin{matrix}  0.01 \\ [-0.02, 0.03] \end{matrix}$                 \\

      $\uparrow \text{LPA}$
                & $\begin{matrix} -0.01 \\ [-0.03, 0.01] \end{matrix}$
                & $\begin{matrix}  0.03 \\ [-0.03, 0.09] \end{matrix}$
                & $\begin{matrix}  0.00 \\ [-0.03, 0.03] \end{matrix}$
                & -
                & $\begin{matrix}  0.00 \\ [-0.02, 0.02] \end{matrix}$                 \\

      $\uparrow \text{SB}$
                & $\begin{matrix} -0.01 \\ [-0.04, 0.02] \end{matrix}$
                & $\begin{matrix}  0.03 \\ [-0.03, 0.09] \end{matrix}$
                & $\begin{matrix} -0.01 \\ [-0.03, 0.02] \end{matrix}$
                & $\begin{matrix}  0.00 \\ [-0.02, 0.02] \end{matrix}$
                & -                                            \\
      \midrule
      \textbf{Within-person level} &&&&& \\
      $\uparrow \text{Sleep}$
                & -
                & $\begin{matrix}  0.04^\ast \\ [ 0.01,  0.07] \end{matrix}$
                & $\begin{matrix} -0.01      \\ [-0.02,  0.01] \end{matrix}$
                & $\begin{matrix} -0.01      \\ [-0.02,  0.00] \end{matrix}$
                & $\begin{matrix}  0.00      \\ [-0.01,  0.01] \end{matrix}$           \\

      $\uparrow \text{Awake in bed}$
                & $\begin{matrix} -0.03^\ast \\ [-0.06, -0.00]  \end{matrix}$
                & -
                & $\begin{matrix} -0.04^\ast \\ [-0.06, -0.00]  \end{matrix}$
                & $\begin{matrix} -0.04^\ast \\ [-0.06, -0.02]  \end{matrix}$
                & $\begin{matrix} -0.03^\ast \\ [-0.06, -0.01]  \end{matrix}$          \\

      $\uparrow \text{MVPA}$
                & $\begin{matrix}  0.01      \\ [-0.01,  0.02]  \end{matrix}$
                & $\begin{matrix}  0.04^\ast \\ [ 0.01,  0.07]  \end{matrix}$
                & -
                & $\begin{matrix}  0.00      \\ [-0.01,  0.01]  \end{matrix}$
                & $\begin{matrix}  0.00      \\ [-0.01,  0.01]  \end{matrix}$          \\

      $\uparrow \text{LPA}$
                & $\begin{matrix}  0.01^\ast \\ [ 0.00,  0.02]  \end{matrix}$
                & $\begin{matrix}  0.05^\ast \\ [ 0.02,  0.07]  \end{matrix}$
                & $\begin{matrix}  0.00      \\ [-0.01,  0.01]  \end{matrix}$
                & -
                & $\begin{matrix}  0.00      \\ [-0.00,  0.01]  \end{matrix}$           \\

      $\uparrow \text{SB}$
                & $\begin{matrix}  0.00      \\ [-0.01,  0.01]  \end{matrix}$
                & $\begin{matrix}  0.04^\ast \\ [ 0.01,  0.07]  \end{matrix}$
                & $\begin{matrix}  0.00      \\ [-0.01,  0.01]  \end{matrix}$
                & $\begin{matrix} -0.01      \\ [-0.01,  0.00]  \end{matrix}$
                & -                                            \\
      \bottomrule
    \end{tabular*}}
    \raggedright{\textit{Notes.} MVPA = moderate-to-vigorous physical activity, LPA = light physical activity, SB = sedentary behaviour.Values are posterior means and 95\% credible intervals. $^\ast$95\% credible intervals not containing 0.}
    \label{tab-sub-out}
\end{table*}

\subsection{Presenting Substitution Analysis Results}
\textbf{\textit{multilevelcoda}} offers a streamlined way of visualising the results
from the \texttt{substitution} models, 
using the \texttt{plot()} method that is built on the well-known \texttt{ggplot2}
package \parencite{ggplot}.
For example, we can graph the between substitution results of sleep by running
\begin{lstlisting}
R> plot(sub_simple, to = "TST", ref = "grandmean", level = "between")
\end{lstlisting}

The estimated differences in stress associated with both the
between- and within substitution results of sleep are in Figure \ref{fig-subplot} 
(some additional parameters had to be set for the figure to be in the format shown,
see supplementary code for details).
Figure \ref{fig-subplot} showed that reallocation from other behaviours 
to time awake in bed was
associated with lower stress level. 
These associations were only significant at the within, but not between-person levels.
\begin{figure*}[!htbp]
\centering
  \caption{Difference in stress for 1-10 minutes of reallocation between 
           time awake in bed and other behaviours.
           TST = total sleep time, 
           WAKE = time awake in bed, 
           MVPA = moderate-to-vigorous physical activity, LPA = light physical activity,
           SB = sedentary behaviour.}
  \includegraphics[width=\textwidth]{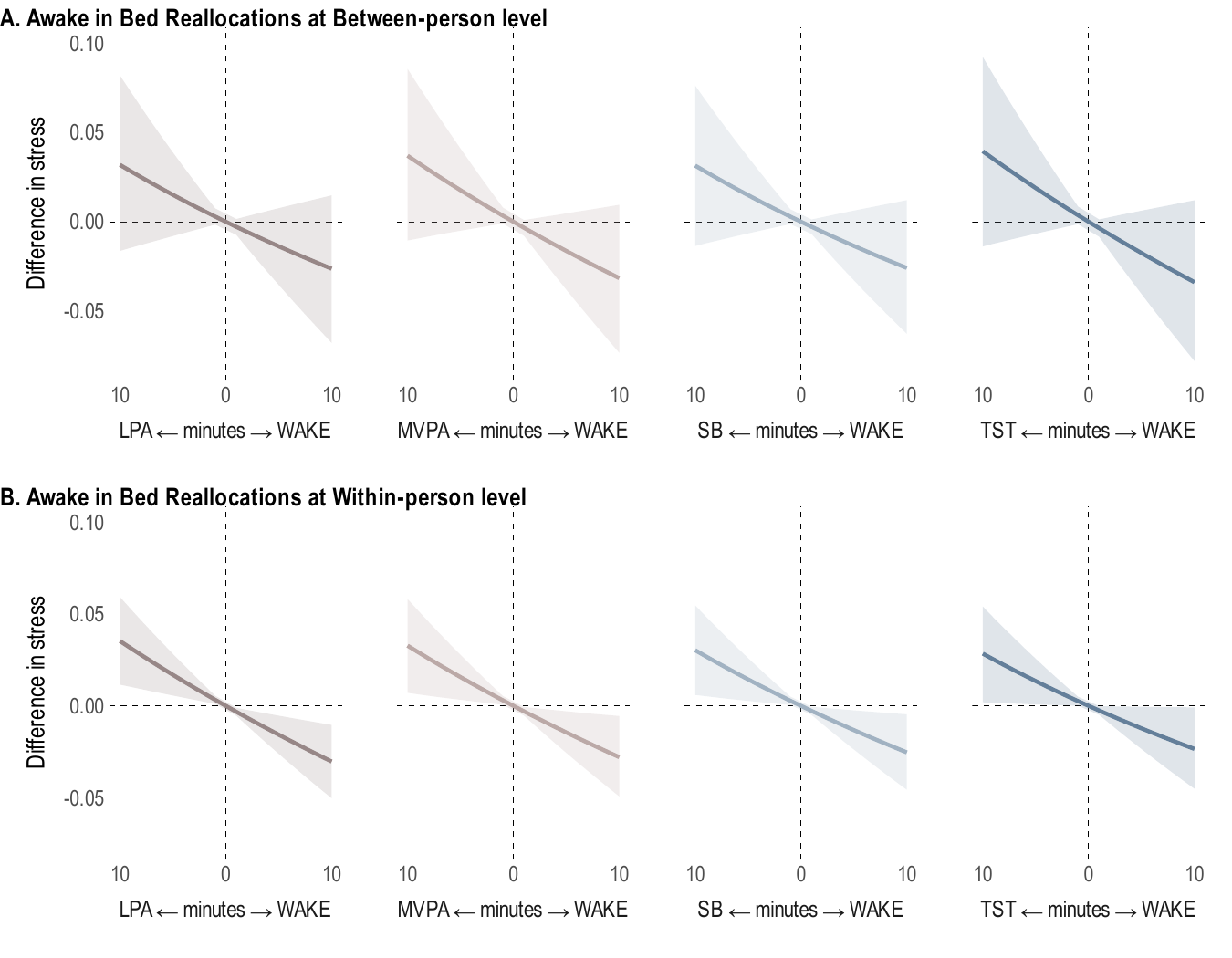}
\label{fig-subplot}
\end{figure*}

\section{Comparison Between Packages}
Many existing \textbf{R} packages implement general functions for compositional data,
however they do not accommodate multilevel data or
provide functions to fit multilevel models with compositional variables and perform post hoc analyses.
The \textbf{R} package \textbf{\textit{multilevelcoda}} stands out by enabling a streamlined workflow for
analysing compositional data in a multilevel framework. Features unique
to \textbf{\textit{multilevelcoda}} are the calculation of multilevel composition and
log-ratios at between- and within-person levels and the capacity
to estimate wide range of (multivariate) multilevel models.
Other features currently exclusive to
\textbf{\textit{multilevelcoda}} is the 
streamlined estimation of pivot coordinates and the multilevel compositional substitution analysis for different types of variability (between and within-person), as well as types of reference composition (grand mean, cluster mean, and user's specified).
Beyond features, another important focus of \textbf{\textit{multilevelcoda}} is on
speed. Models using \texttt{brmcoda()} are fitted using package \textbf{\textit{brms}}, which
generally require more time and computational resources than package \textbf{\textit{lme4}}.
Given the complexity of the substitution analysis, \texttt{substitution()}
supports parallel execution via package \textbf{\textit{foreach}} \parencite{foreach} and
\textbf{\textit{doFuture}} \parencite{doFuture}, which enables models to run faster in
shorter walltime.
A comparison across packages for working with compositional data is
provided in Table \ref{tab-compare}.

\begin{table*}[!htbp]
\caption{Comparison across packages for compositional data. 
         \textit{Notes.} ilr = isometric log-ratio, alr = additive log-ratio, clr = centered log-ratio.
         \textsuperscript{*}Models with compositional outcomes can include compositional predictors.
         \textsuperscript{\textdagger}Only available for Bayesian models.
}
\centering
\resizebox{\textwidth}{!}{\begin{tabular}{llllll}
  \toprule
  & \textbf{\textit{multilevelcoda}}                & \textbf{\textit{compositions}}  & \textbf{\textit{Compositional}} & \textbf{\textit{compositions}} & \textbf{\textit{robCompositions}} \\[4pt]
  \midrule
  \noalign{\vskip 1mm}
  \textbf{Basic functions for composition} \\[4pt]
  Composition                                                   
  & Yes                                 & Yes                 & Yes                 & Yes                 & No            \\
  Multilevel composition                                        
  & Yes                                 & No                  & No                  & No                  & No            \\
  Logratio transformation                                       
  & ilr, alr, clr                       & ilr, alr, clr       & ilr, alr            & No                  & ilr, alr, clr \\
  Multilevel logratios                                   
  & Yes                                 & No                  & No                  & No                  & No            \\
  Missing value and zero imputation                             
  & No                                  & Yes                 & Yes                 & Yes                 & Yes           \\
  Outlier detection                                             
  & No                                  & Yes                 & No                  & No                  & Yes           \\[4pt]
  \textbf{Frequentist models} \\[4pt]
  Single-level with compositional predictors                    
  & No                                  & No                  & Yes                 & No                  & No            \\
  Single-level with compositional outcomes                      
  & No                                  & No                  & Yes                 & No                  & No            \\
  Multilevel with compositional predictors                      
  & No                                  & No                  & No                  & No                  & No            \\
  Multilevel with compositional outcomes                        
  & No                                  & No                  & No                  & No                  & No            \\[4pt]
  \textbf{Bayesian models} \\[4pt]
  Single-level with compositional predictors                    
  & Yes                                 & No                  & No                  & No                  & No            \\
  Single-level with compositional outcomes\textsuperscript{*}
  & Yes                                 & No                  & No                  & No                  & No            \\
  Multilevel with compositional predictors                      
  & Yes                                 & No                  & No                  & No                  & No            \\
  Multilevel with compositional outcomes\textsuperscript{*}                                 
  & Yes                                 & No                  & No                  & No                  & No            \\[4pt]
  \textbf{Pivot coordinate estimation} \\[4pt]
  Single-level with compositional predictors                    
  & Yes\textsuperscript{\textdagger}    & No                  & No                  & No                  & No            \\
  Single-level with compositional outcomes\textsuperscript{*}
  & No                                 & No                  & No                  & No                  & No            \\
  Multilevel with compositional predictors                      
  & Yes\textsuperscript{\textdagger}    & No                  & No                  & No                  & No            \\
  Multilevel with compositional outcomes\textsuperscript{*}                                 
  & No                                 & No                  & No                  & No                  & No            \\[4pt]
  \textbf{Compositional substitution analysis} \\[4pt]
  Simple single-level                                                  
  & Yes\textsuperscript{\textdagger} & No                  & No                  & No                  & No            \\
  Simple multilevel                                                    
  & Yes\textsuperscript{\textdagger} & No                  & No                  & No                  & No            \\
  Average single-level                                         
  & Yes\textsuperscript{\textdagger} & No                  & No                  & No                  & No            \\
  Average multilevel                                           
  & Yes\textsuperscript{\textdagger} & No                  & No                  & No                  & No            \\[4pt]
  \bottomrule
\end{tabular}} \\
\label{tab-compare}
\end{table*}

\section{Future extension}

In this article, we introduced the analysis of multilevel compositional data in a
Bayesian framework using the our \textbf{R}
package \textbf{\textit{multilevelcoda}}. In the current limited landscape for modelling
multilevel compositional data, \textbf{\textit{multilevelcoda}} is a contribution
that integrates three methods: compositional data analysis, multilevel
modelling, and Bayesian inference into one, open-source program. The
implementation of \textbf{\textit{multilevelcoda}} enables a streamlined and
efficient workflow from dealing with raw multilevel compositional data,
estimating models, and presenting final results, making the analysis of
multilevel compositional data faster and more accessible. To the best of
our knowledge, this is the first statistical package provides tools for
estimating multilevel isotemporal compositional substitution analysis at
between and within-person levels. As this method provides a unique
opportunity to gain novel insights in the fields of epidemiology and
psychology, such as the integrated and interactive effects of sleep-wake
behaviours on health outcomes, \textbf{\textit{multilevelcoda}} may be
particularly useful for intensive longitudinal studies in this
landscape. The support for optional parallel execution further promotes
an efficient and powerful performance, enabling the estimation of
complex models with less walltime.

\textbf{\textit{multilevelcoda}} is under active development. 
A current priority is to 
support various outcome families (e.g., multivariate) in the substitution analysis. 
We also plan on integrating features to deal with
missing data, zeros, and outliers from existing packages to enable more
streamlined workflow. Functions to estimate the marginal
means for Bayesian multivariate multilevel models with compositional
outcomes when integrating out group-level effects if desired will also
be added. These extensions will be made available on the
developmental version on GitHub before releasing on CRAN, along with
vignettes to demonstrate new functionality. Interested users are welcome
to follow the \href{https://github.com/florale/multilevelcoda}{GitHub
version of \textbf{\textit{multilevelcoda}}} and provide feedback for future development of the package.

\printbibliography
\end{document}